# Blueprints of Trust: AI System Cards for End-to-End Transparency and Governance

Huzaifa Sidhpurwala
huzaifas@redhat.com

Emily Fox
efox@redhat.com

Garth Mollett
gmollett@redhat.com

Florencio Cano Gabarda
fcanogab@redhat.com

Roman Zhukov
rzhukov@redhat.com

## Abstract

This paper introduces the **Hazard-Aware System Card (HASC)**, a novel framework designed to enhance transparency and accountability in the development and deployment of AI systems. The HASC builds upon existing model card and system card concepts by integrating a comprehensive, dynamic record of an AI system's security and safety posture. The framework proposes a standardized system of identifiers, including a novel **AI Safety Hazard (ASH) ID**, to complement existing security identifiers like CVEs, allowing for clear and consistent communication of fixed flaws. By providing a single, accessible source of truth, the HASC empowers developers and stakeholders to make more informed decisions about AI system safety throughout its lifecycle. Ultimately, we also compare our proposed AI system cards with the ISO/IEC 42001:2023 standard and discuss how they can be used to complement each other, providing greater transparency and accountability for AI systems.

We encourage readers of this paper to review our our previous work[1], as several concepts in this paper are built upon proposals and concepts described there.

---

[1] https://arxiv.org/pdf/2411.12275

# Introduction

Generative AI has evolved from an experimental curiosity to a pervasive general-purpose technology in just a few years, reshaping workflows, products, and cultural expectations across every sector of the economy. Ultra-large multimodal foundation models—capable of synthesizing text, images, code, audio, and even video—now anchor consumer chatbots, enterprise copilots, creative suites, and developer tools, with daily usage measured in the hundreds of millions. Businesses cite productivity gains, accelerated innovation cycles, and new revenue streams as key benefits, while individuals increasingly rely on AI for tasks ranging from language translation and personalized education to content creation and software prototyping. At the same time, the technology's democratizing reach, significantly propelled by the community AI ecosystem growth, raises urgent questions about data provenance, intellectual property rights, labor displacement, and the amplification of misinformation, prompting an unprecedented wave of regulatory activity, from the EU AI Act to sector-specific guidelines in healthcare and finance.

Market Research estimates that the global artificial-intelligence market is valued at USD 638.23 billion in 2024 and projected to reach USD 757.58 billion in 2025, expanding at a compound annual growth rate of roughly 19 percent to an impressive USD 3.68 trillion by 2034.[2] This sharp, sustained trajectory highlights how AI's perceived capacity to boost operational efficiency and data-driven decision making is accelerating its adoption across virtually every core industry, providing both the economic context and the technological imperative for the present study.

# From AI models to AI systems

Since mid-2023, the Hugging Face Hub[3] has doubled in size roughly every 10 months, crossing one million models in late 2024 and topping 1.7 million by mid-2025. Yet usage remains highly skewed: foundational checkpoints such as BERT, a bidirectional AI language representation model, still draw tens of millions of monthly downloads, while state-of-the-art Mixture of Experts (MoE) giants attract orders-of-magnitude fewer pulls. The composition of the Hub is also shifting.Natural Language Processing (NLP) models now represent only about half of new uploads as use case specific models, such as. vision, audio, and code, surge. Combined with petabyte-scale storage and incomplete documentation for over half the repositories, these trends illustrate both the vitality of the community AI ecosystem and the escalating need for system-level governance artifacts. Faster–than–ever community contributions and shared advancements, simultaneously highlights the escalating need for standardized system–level governance artifacts like AI System Cards, which can be collaboratively defined and adopted by the very communities driving this growth.

---

[2] https://www.precedenceresearch.com/artificial-intelligence-market?utm_source=chatgpt.com
[3] https://huggingface.co/models

While advanced AI models like GPT4o demonstrate remarkable capabilities, they have limited practical use in isolation. The model itself is often likened to an engine: a powerful component that by itself *"can't go anywhere"* without a chassis, wheels, steering, and other parts to form a complete car. Indeed, real-world deployments show that *"only a small fraction"* of an ML system is the model code, whereas *"the required surrounding infrastructure is vast and complex"*[4]. In other words, the majority of an AI system consists of the supporting components that allow the model to function usefully, safely, and reliably. Thisproverbial vehicle makes the engine's power accessible.

To be practically useful **and** safe, an AI model must be deployed as part of a complete system within a secured lifecycle encompassing interfaces, integration, infrastructure, and oversight. For example, models like GPT4o only become valuable when embedded behind user-friendly interfaces,for example, ChatGPT in this case, and connected to other tools and data via APIs. Such integration provides the "plumbing" that feeds the model with the right information and channels its outputs into real-world actions like populating a dashboard or sending an email. Equally important are safety guardrails and monitoring: powerful models *"can be unpredictable, biased, or produce harmful outputs if left unchecked."* Defining boundaries and filters are critical to keep the AI's behavior within acceptable limits. Only by integrating the model into a full system of interfaces, infrastructure, guardrails, and governance can we harness its power effectively and securely.

## Growth of the AI ecosystem -infrastructure, tooling and applications

Over the past two years, the rise of widely-available (and free of charge) tools across the AI, ML, and data ecosystem has not only broadened access but also fundamentally lowered the barrier to building advanced AI systems. As highlighted in the 2024 MAD Landscape,[5] tools such as Hugging Face, Jupyter, Streamlit, Metabase, and Airbyte now cover nearly every stage of the AI development lifecycle—from data ingestion and transformation to model training, deployment, and monitoring. This accessibility allows individual developers, startups, and research teams to assemble production-grade AI stacks without the prohibitive licensing costs and vendor lock-in associated with proprietary platforms.

---

[4] https://proceedings.neurips.cc/paper_files/paper/2015/file/86df7dcfd896fcaf2674f757a2463eba-Paper.pdf
[5] https://mad.firstmark.com/

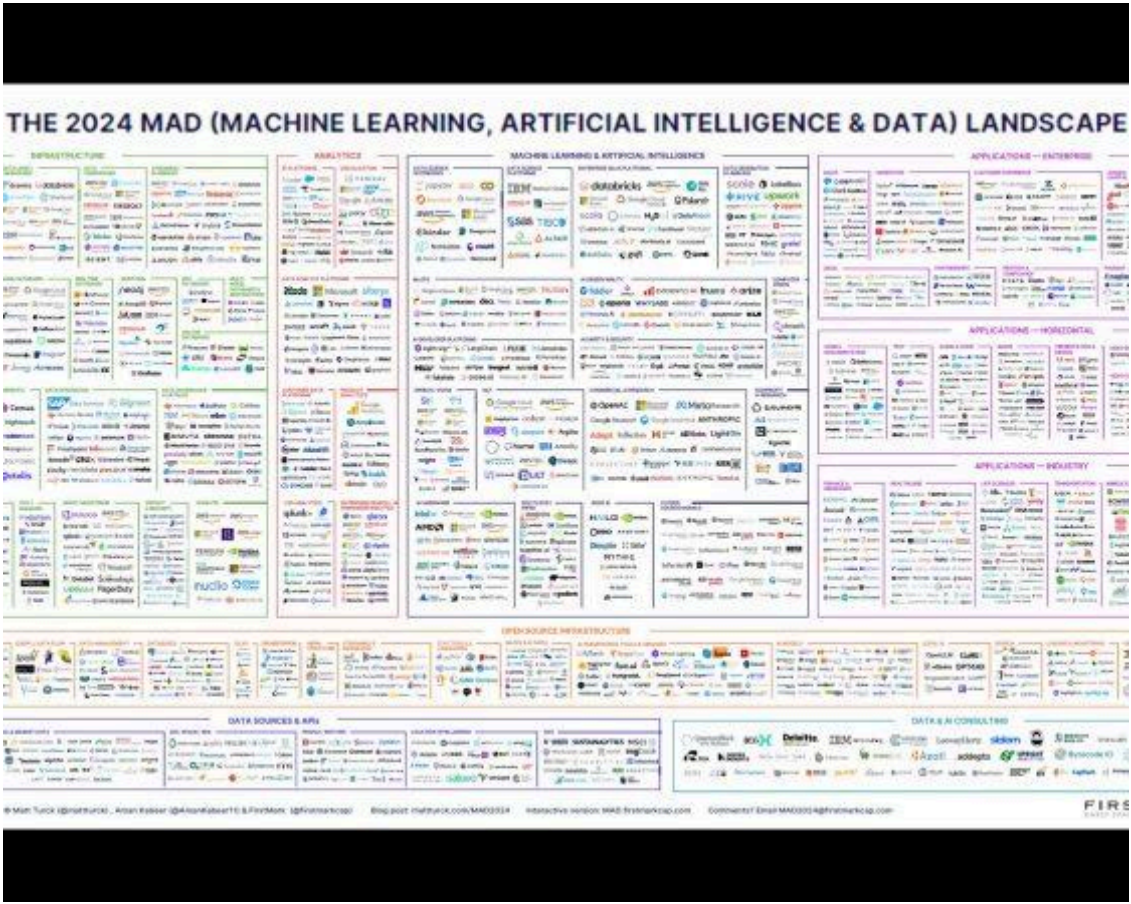

The above image is reproduced from:: https://mad.firstmark.com/

The open source model fosters rapid experimentation. Frameworks like PyTorch and TensorFlow, coupled with model hubs such as Hugging Face, let developers integrate state-of-the-art architectures with minimal setup. Data engineering and integration platforms like Meltano make it straightforward to connect heterogeneous data sources, streamlining the pipeline from raw data to model-ready datasets. Likewise, tools like Streamlit and Metabase allow developers to wrap models in user-friendly interfaces and dashboards quickly, enabling fast iteration cycles and stakeholder feedback loops.

Crucially, these tools promote interoperability and modularity (building on cloud native concepts), meaning AI systems can be built by combining best-in-class components rather than reinventing the wheel. Developers can source pre-trained models, plug them into existing data pipelines, add MLOps layers for version control and monitoring, and ship a working prototype in days rather than months. This modular ecosystem is why the pace of AI system development has accelerated dramatically; smaller teams can now deliver capabilities that previously required enterprise-scale resources.

Ultimately, the prevalence of these platforms has democratized AI system development. By removing cost and complexity barriers, they have enabled a more diverse range of innovators—from independent researchers to mid-size enterprises—to contribute to and benefit from the AI revolution.

## Models as Systems and the need for AI System Cards

As AI system development becomes accessible to a broader range of innovators due to such platforms, the reliance on AI Model cards to effectively convey pertinent information to those developers, researchers, and organizations has revealed key gaps critical to allowing entities to assess their risk in using models and systems that integrate them. Where model cards provide the core documentation of individual machine learning models, integrators and adopters need more transparency to effectively identify and manage risks associated with the model in the context of the system it's being used. This information is essential not only in the risk determination, but also supports hazards and vulnerability management efforts, which heavily rely on understanding the system to triage correctly and assess impact.

### AI model cards and their extensions

Model cards, initially proposed in 2018,[6] are not new to the AI ecosystem. They are short, human-readable docs that accompany a Machine Learning (ML) model to explain **what it's for** and **how to use it responsibly**. Currently, model cards are widely used in the AI industry, mainly by model makers.[7]

In our earlier work,[8] we proposed several extensions to the standard model card. More specifically, the paper proposes to **standardize and extend model cards** so they're comparable, actionable, and interoperable across tooling. Concretely, it recommends:

- Creating a specification with a consistent minimum field set. This means standardizing the model card across model creators.
- Adding an "Intent and Scope" of the model. The Intent talks about users and the use cases of the model, while the scope purpose of Scope is to exclude known issues that the Model producer has no intent or ability to resolve.
- Add data governance and pedigree information, such as data provenance. What data was used to train and fine-tune the model, from where was the data obtained, how was the data cleaned, what criteria was used to clean the data etc.

---

[6] https://arxiv.org/abs/1810.03993
[7] https://huggingface.co/ibm-granite/granite-3.3-8b-instruct
[8] https://arxiv.org/pdf/2411.12275

- We spoke about adding security and safety information to the model cards. We proposed the HeX format, similar to VeX[9], for keeping track of safety hazards reported and fixed in the model.
- Lastly, we spoke about AI SBOM[10] and linking them to the model card, along with possible integration with OCI.

In the absence of these changes, there remains an outstanding need for this information to be conveyed. With the increased integration of models into systems, AI system developers are well-positioned to record and convey these critical data points, akin to those we see within the software ecosystem. By shifting beyond a unique model-focus to the actual use of these models within an operational environment and business context, we transparently bring the AI ecosystem and the existing technology ecosystem closer to achieving operational maturity together.

The concept of system cards was formalized by Meta[11] as "**System-Level Transparency**," arguing that risk lives at the *system* boundary (data → model → product) rather than the model alone, further implying that the use of model cards alone are insufficient when operating AI systems in enterprise environments. This shift aligns with guidance from **NIST AI RMF**[12] to document end-to-end risks and mitigations as well as requirements detailed in **ISO/IEC 42001** and the **EU AI Act**.[13]

## Why Model Details and Data Provenance Matter

The Model and Data details category of content is closest to the model card with a notable distinction– model cards that talk *about* the model. Model cards are designed to explain the how and performance of the model most relevant to a machine learning and data science audience. The Model and Data details conveyed in an AI System Card go beyond this by providing additional and comprehensive information about the data used for training and fine tuning. That is where the concept of Data Provenance becomes essential in the entire AI implementation lifecycle.

Organizations have already heavily relied on data for critical decisions, operations, and risk management, demanding stringent verification of its origin, quality, and authorized use. The rapid adoption of AI brings the entire supply-chain trust and accountability concerns to a dramatically new scale. To tackle these challenges it will be important to apply the new standards like data provenance. Data Provenance is a consistent framework used to track the origin, movement, integrity, and characteristics of data, essential for understanding a data set's origin, quality, and intended use. Data Provenance is critical because it establishes a verifiable audit trail for data in the standardized machine-readable format, encompassing

---

[9] https://www.cisa.gov/sites/default/files/publications/VEX_Use_Cases_Aprill2022.pdf
[10] https://owasp.org/www-project-aibom/
[11] https://ai.meta.com/research/publications/system-level-transparency-of-machine-learning
[12] https://www.nist.gov/itl/ai-risk-management-framework
[13] https://artificialintelligenceact.eu/

its origin, transformations, and usage throughout its lifecycle. This granular lineage is essential for:
- **Verifiable Lineage with Contextual Metadata**: Provenance provides a comprehensive record of data's journey, including timestamps, actors, processes, and systems involved at each stage. This contextual metadata (e.g., sensor calibration data, algorithm version, data cleaning scripts) allows for a deep understanding of data quality and reliability.
- **Built-in Integrity Checks and Audit-Readiness**: By recording cryptographic hashes or digital signatures at various points in the data supply chain, provenance enables robust integrity checks. This ensures data has not been tampered with and facilitates efficient auditing by providing an immutable record for regulatory compliance (e.g., GDPR, HIPAA) and internal governance.
- **Centralized Access for Evaluating Reliability and Compliance**: A centralized provenance repository allows for systematic evaluation of data's trustworthiness. This enables data consumers to assess the data's fitness for purpose, understand potential biases, and verify adherence to organizational policies and regulatory mandates.
- **Supports Operational Transparency and Explainable AI (XAI)**: In complex data ecosystems, provenance offers transparency into data flows, aiding in debugging and performance optimization. For AI/ML models, provenance links model outputs back to their input data and training processes, contributing to explainability by demonstrating how specific data points influenced model predictions, which is crucial for ethical AI and accountability.

While the standardization work around the Data Provenance framework and its technical details (like machine-readable formats) is not completed yet, a number of high value use-cases are already demanded by the industry. Various roles across data science, engineering, legal, and procurement would need to rely on data provenance to ensure trust, compliance, security and usability of data. For example, data scientists and engineers need to validate data rights, freshness, and privacy before building or sharing models. Product managers require transparency in data generation, source quality, and downstream use constraints to select or build effective AI solutions. Procurement, legal and governance specialists depend on provenance to assess legality, redundancy, and data quality. Across all cases, data provenance is critical to reduce risk, enable compliant data use, and maintain confidence in AI and analytics outcomes.

The model and data details such as Data Provenance can support adopters in defining the potential quality of the AI system's output, building trust through self-reporting breadth and depth of its dataset.  In simple terms, the model card provides an overview., and the model and data details of the system card provides you with a manual of understanding on how the model was incorporated into the final system. When it comes to risk management, this information can support discovery of hazard and vulnerability sources, which is essential for driving mitigation and remediation; the core to a successful risk management program and necessary to demonstrate compliance with standards for quality, accuracy, and security such as those detailed by the EU AI Act.

## What makes AI system cards useful

AI System cards are generally divided into four categories of content, each with its own relevance to existing organizational processes for acquiring, adopting, and integrating technology and the corresponding risk these activities and technologies introduce.

The System Overview and Intended Use category closely aligns with content conveyed on the model card itself, the details and intended use. We can and should consider the model card as a sub-section or linked external content to the overall AI system card for completeness in consideration. For entities seeking to make use of AI systems, this section in the card can inform adopters what is out of scope or prohibited in use - allowing them to understand if the system functionally aligns with the need and understand where additional liability or risk can transpire from inappropriate use. It also defines the operational boundaries of the system, essential for understanding whether use can result in a safety hazard likelihood of occurrence.

For organizations considering AI risk frameworks like NIST and ISO/IEC 42001/23894, or in scope of legislation like the EU AI Act, this category provides governance and contextual mapping necessary to fulfill several requirements that contribute to risk classification, defining policies and controls, and make risk of inappropriate use explicit.

The evaluation and performance metrics category details how the AI system's performance was measured, building on the performance information of the model itself, and the corresponding results of those evaluations. For adopters, this sets realistic expectations for the AI system's reliability and accuracy, further building trust in use. It can provide an indicator of effectiveness, curtailing the extensive needs to babysit an unreliable model that can negatively create more risk and harm in businesses using it. This does not remove the need for a human in the loop for decision making when using AI systems, but can provide a performance threshold essential for organizational process integration to assess where and at what level that should occur. These evaluation metrics directly contribute to understanding the AI system's inherent risk of failure. This information is crucial to industries such as financial services, banking, healthcare, citizen services, and governments. If an AI system's evaluation metrics indicate low accuracy on key topics, the probability of a hazard occurring from its use in those topical areas is increased and informs the quantitative analysis required to understand not only the likelihood of that hazard realization, but also the magnitude of the potential harm that may result.

The last category of content in the AI system card is perhaps the most actionable for adopters because it directly conveys known weaknesses about the AI system. Limitations and known biases transparently inform the potential of a hallucination by a model, which can result from knowledge scope or freshness of the knowledge base. This allows organizations to identify compensating controls to reduce the risk introduced by the AI system. This can be building additional guardrails or introducing key checkpoints for a human-in-the-loop to validate and verifying output before action is taken. In an industry seeing rapid development and change, the freshness of the knowledge base is essential to understanding if the output is based on a topic that must be flagged for review to understand its recency. The content in this category directly guides and informs any impact assessment being conducted, such as those required by ISO 42001 and is a key point in the transparency pillar of the EU AI Act.

These four categories of content, located in the AI system card, directly contribute to an organization's effective risk management for the use of AI in its infrastructure and business processes.

## AI system cards and their use in the industry

Currently, system cards are being published by several major frontier model creators:

- **OpenAI** – System cards for frontier models and features, for example, GPT-5.[14]. They describe capabilities and limits across modalities, pre andpost-deployment safety tests, and *third-party* evaluations, such as METR and Apollo, for autonomy and persuasion risks.
- **Anthropic** – **Claude** - System cards like Claude Opus 4.1.[15] with Anthropic's Responsible Scaling Policy (RSP)-aligned safety testing, prompt-injection and computer-use risks, and extensive red-team evaluation sections. Recent cards and addenda track new reasoning and coding capabilities and the associated hazards.
- **Meta** – A broad library of 22 system cards explaining major AI systems behind Facebook andInstagram, ranking information, such as,reeds, reels, explore, and search, plus later cards for generative AI features. The cards aim to be readable for general users and show signals, controls, and personalization choices.

Other model creators leverage alternative structured formats to convey similar data found within a system card:
- **Microsoft** – Publishes Transparency Notes[16] for Copilot products that are functionally system-card equivalents, what data they use, safety boundaries, failure modes, and deployment controls. The notes now extend across services and developer surfaces.
- **Google** – Emphasizes model cards, such as Gemini andGemma,[17] and system-level safety docs and playbooks for developers, to include policies and evaluations.

---

[14] https://cdn.openai.com/gpt-5-system-card.pdf
[15] https://assets.anthropic.com/m/4c024b86c698d3d4/original/Claude-4-1-System-Card.pdf
[16] https://learn.microsoft.com/en-us/copilot/microsoft-365/microsoft-365-copilot-transparency-note

Lack of standardization on the content structure and the format for conveying this critical information can and will contribute to confusion by adopters of AI systems looking to compare products and projects against organizational needs and risk tolerances. Regardless of the structure and overall content, they all heavily rely on human-generated content to convey these concepts and statistics, which should begin to be shifted to machine-readable content for automated risk management and policy creation.

## Unaddressed elements in the currently published System Cards

Several of our proposed extensions to model cards are still outstanding, with system cards highlighting the continuing gap in making model consumption and deployment a challenge for adopters considering the overall risk use of a model presents to their systems and business.

Interoperability and comparability is essential for both pre and post operational decisions of models. Across organizations and companies, the sections, metrics, and risk taxonomies differ, making cross-vendor comparison by AI system adopters exceedingly difficult.

Frameworks like **CLeAR** specifically call for *Comparable, Legible, Actionable, Robust* documentation, however many cards aren't yet comparable or machine-testable.[18] Most cards are PDFs and web pages, with very few exposing a structured schema for tools to ingest and for use in policy enforcement and audits. NIST AI RMF encourages operationalization, but there's no de-facto schema in place today.

With minimal or no information about the provenance and pedigree of the data used to train the model or the pipeline used to clean the data before training, consumers and adopters of AI systems cannot independently verify and validate any claims purported by the producer of an AI system card.

A few cards exist as living documents that log field failures, drift, and mitigations over time—despite NIST's emphasis on lifecycle risk management, which goes beyond the point of creation to articulate everything that has transpired to the state at which the AI system exists today. These make no mention of any security or safety incidents remediated in the current release(s) of the system. The dynamic nature of models and their premise for inherent bias demonstrates that the history of the model and the system built around it are equally important in understanding the risk a system presents. Analogous to software's changelogs, CVEs, and SBOMs, AI needs *living* system cards that record provenance, updates, and remediated hazards across releases.Without them, stakeholders can't contextualize risk or verify claims.

---

[17] https://storage.googleapis.com/model-cards/documents/gemini-2-flash.pdf

[18] https://shorensteincenter.org/clear-documentation-framework-ai-transparency-recommendations-practitioners-context-policymakers

While the current industry and regulatory focus remains on LLMs, the proliferation of powerful AI systems interfacing directly with humanity highlights the volume of poorly documented, yet significant hidden risks to daily life. Meta's library is exemplary for recommenders, but across the industry, many safety documents focus on LLMs, while other high-impact AI systems, such as ads ranking, trust and safety automation, remain unevenly documented,[19] especially related to security and safety details. This spotlight effect needs to be addressed. Developers and innovators have moved beyond the models to the practical use of these in day-to-day activities, and so too must the documentation, controls, and regulatory requirements. It is imperative to ensure such impactful systems meet the same level of scrutiny that the LLMs they're built upon are required to undergo.

Models are only one part of an AI system. The end-to-end risk profile is shaped just as much by the serving layer, theAPI gateway,inference server, and autoscaler, along with caching, such as KV and prompt caches, and also orchestration, for example,  Kubernetes and OpenShift, data connectors and vector stores, guardrails and policy engines, and observability. Most system cards narrowly focus on the backend model, omitting these non-model components and their security and safety controls like authN/Z, isolation, patch levels, logging and retention, rate limits. Documenting them delivers concrete value. It reveals the attack surface and data flows, explains failure modes, enables reproducibility and incident response like trace a leak to cache retention rather than the model, and ties mitigations to versions for governance and audits. For example, a RAG assistant's risk often hinges on vector-DB ACLs, connector scopes, and cache policies more than the base model itself. Without these details, stakeholders cannot accurately assess system risk.

## Principles behind open and transparent system cards

In this paper, we present open system cards: standardized, machine-readable dossiers for AI systems that go beyond capability benchmarks to include security flaws, safety hazards, and their remediations across versions. Rather than asserting that one model is "better" than another, system cards enable side-by-side, evidence-based comparison—covering capability metrics under stated conditions, data provenance and governance, deployment and guardrail configurations, known limitations, and a versioned record of hazards and fixes. This is so stakeholders can assess suitability for a given use case.

---

[19] https://transparency.meta.com/features/explaining-ranking/fb-feed-recommendations/

We present the Hazard Aware System card framework (HASC), which is a **living document** designed to be the central hub for identifying, monitoring, and responding to AI system hazards in real-time. It is designed to serve as the primary vehicle for achieving transparent and open system safety of AI, providing internal stakeholders with actionable intelligence and external stakeholders with clear, trustworthy information about system performance and incident response. By adopting this framework, model creators can move from a reactive to a proactive and transparent safety posture, building trust and demonstrating industry leadership.

The remainder of this paper discusses this new framework, examines some significant challenges and strategies, and provides real-life examples of how these open system cards can be useful and applied effectively.

## The Hazard-Aware System Card (HASC)

The HASC expands the traditional AI System Card with new, dynamic sections focused on safety and response. It is designed to be integrated directly into the AI development lifecycle and operational monitoring tools. Equally important, the HASC is designed to plug into the AI development lifecycle and the ops toolchain. It is meant to be a machine-readable entity, generated during the model training and AI system build process, for consumption by external entities wanting to use the systems.

### Components of HASC

The type and amount of information made available in a system card depends on the amount of transparency the vendor would like to provide. In this section, we present a few components that we feel are essential and need to be included for the HASC to make sense. We also include optional components that vendors can add to deepen assurance without forcing disclosure of sensitive intellectual property. These optional items can be phased in, based on the system's risk level and the organization's maturity. It is worth mentioning that some content is manually generated with less probability of change, whereas others can be automatically generated.

Essential components

**System blueprint**

The System Blueprint section serves as the foundational, static component of the HASC, providing a comprehensive technical overview of the AI system. See the following, more detailed, breakdown of the components that would be included in the System Blueprint:

- System architecture: This is a high-level overview of the AI system, detailing its AI and non-AI components, data flows, and where the core models perform their inference. It includes diagrams showing how different microservices or modules interact, from data ingestion to user output.
- Model and data provenance: This section provides links to the "AI Bill of Materials" (AI SBOMs) to show which models are used by the system. It should list the specific core models used, including their version numbers, for example, `model-name-v2.1`, as well as links to the additional metadata such as verifiable data provenance
- Intent and scope of the system: This is inline with the concepts explored in our previous paper[20].

**Proactive hazard analysis**

This section of the HASC is designed to proactively identify and document potential risks associated with the AI system before deployment. It is a foundational component of the framework. Here is a breakdown of its key components:

- Hazard Log: The Hazard Log is a structured list of potential harms that have been identified through various pre-deployment safety exercises such as red teaming, threat modeling, and ethical reviews. Examples of these potential harms include severe bias, the generation of unsafe content, privacy violations, and manipulation risks. Ideally the Hazard Log should only reference HeXs regardless. We discuss HeX in more detail later in the paper.
- Hazard probability score: Risk is all about context, and without the operational environment or use, you cannot ascertain potential impact. If the hazard is identified in the system, its occurrence and interference can be expected . Therefore, the probability of the hazard occurring given an input value is the most appropriate metric here.
- Embedded guardrails: This component provides a clear description of the specific technical and policy safeguards that have been designed to mitigate each identified hazard. It serves as a record of the safety measures implemented to protect against the risks outlined in the Hazard Log.

**Incident response and hazard remediation**

Our previous paper discussed the distinction between AI security flaws and safety hazards in detail. The AI system card should contain a section that includes all of the security and safety issues fixed in the current, specified version of the system. If a dynamic section is not feasible, then at minimum, a link to such a publicly available document should be provided.

---

[20] https://arxiv.org/abs/2411.12275

This includes any CVEs[21] fixed to remediate security issues and any hazard numbers which were assigned and fixed to remediate safety issues. Only issues related to components listed in the system card should be mentioned. We propose hazard numbers or identifiers in the section below, since they are applicable to both models as well as systems.

Optional components

Several optional components can be included in the AI system card, depending on the type of system in question, and how transparent the vendors can be. These include, but are not limited to the following:

- The inference engine used in the model's backend.
- The agentic architecture used in the system.
- The hosting platform.
- Open source components used to make up the system, for example, vector databases and front end tools.

## Hazard Identifiers and HeX

In our previous work, we described in detail the difference between AI security and safety issues. In the case of AI security, the fact is that the security ecosystem is well researched and organized, with proper assignment, tracking, and management of vulnerabilities which each one has an identifier called CVE[22] ID. We also proposed a possible system for doing the same with safety hazards which involved a separate committee for managing such safety identifiers. In the absence of such a centralized system, we propose assignment of IDs by the same organizations that build these AI systems. This is not an ideal solution, and such a system can quickly lead to confusion, overlapping IDs and general misinformation about the nature of the hazard, its affected components, and its fixes. However, the authors feel that in the absence of a centralized solution this arrangement is better than having none at all. Authors of this paper strongly suggest engagement with public AI forums to centralize the management and governance of such a system to deal with hazard identifiers.

In either case, we propose an identifier of the format:
**<common identifier>-<year>-<number>.** For example, ASH-2025-0023: AI Safety Hazard (ASH), the year in which it was discovered, and the running number of the hazard identifier within that year.

We also propose to use this hazard identifier in the AI system card as well as HeX data, as discussed in our previous work.

---

[21] https://www.cve.org/Media/News/item/blog/2024/07/09/CVE-and-AIrelated-Vulnerabilities
[22] https://www.cve.org/

## Generation and consumption of the AI system cards

AI system cards are most effective when treated as code and generated automatically from the software delivery pipeline. A JSON Schema[23] serves as the contract that details what must be captured: metadata, intent or scope, model and guardrails versions, data provenance, evaluation results, hazards and mitigations, governance contacts, and references. During build and deploy, CI jobs auto-populate the card from authoritative sources: model registries for model/version and evaluations, IaC and Kubernetes/OpenShift manifests for hosting and topology, guardrails/moderation configs for policy versions and thresholds, data catalogs or ML-BOMs for training/augmentation lineage, and issue trackers for hazard records and status. The resulting JSON instance is validated against the schema, signed with a component such as in-toto/SLSA attestations, versioned with the release tag, and rendered to Markdown/HTML for human consumption. This automation minimizes manual burden, keeps cards current by construction, and yields diffable, auditable artifacts suitable for post-mortems and external reviews.

Since the card is machine-readable, it can be consumed by multiple control points without bespoke integrations. Organizations' product security policies can enforce "release gates" with policy-as-code, for example, blocking a production deploy if the card lacks a security contact, if hazards exist without mitigation status, or if the guardrails version regresses. Risk and compliance teams can aggregate cards into an inventory that highlights stale entries, missing references, or systems whose data provenance does not meet organizational standards. Operations teams can wire incident runbooks to the card's "Known hazards → Mitigations," accelerating containment, for example, toggling a stricter prompt policy or rolling back a model version documented in the card. Vendor-management and procurement can export a concise assurance profile from the same JSON to answer due-diligence questionnaires, map to frameworks, for example, NIST AI RMF-aligned controls, and demonstrate supply-chain transparency via SBOM/ML-BOM links.

Standardized fields enable cross-team and cross-vendor comparison: two systems produced by different groups can still be scored by the same policy engine and visualized on a common dashboard. System cards capture both technical and governance signals, making them a source of durable evidence for audits. For example, model changes tied to dates, rationale, and eval deltas, guardrails versions and thresholds, and provenance updates. Over time, organizations can analyze card histories to identify systemic risks and trends—say, recurrent jailbreak classes or data-quality regressions, and prioritize structural fixes rather than one-off patches.

---

[23] https://github.com/RedHatProductSecurity/ai-system-card

## Major challenges and mitigation strategies

AI system makers face clear challenges in adopting the AI system card paradigm. Some organizations have navigated this by issuing partial cards, including certain components but omitting others. Conversely, some completely ignore the paradigm, releasing only enough information to stay competitive.

**Challenge: Balancing competitiveness and transparency**

This challenge represents the fundamental tension between the desire for comprehensive AI transparency, as championed by frameworks like the HASC, and the competitive realities of the AI industry. While transparency is increasingly recognized as crucial for trust, regulation, and safety, companies often have strong incentives to guard proprietary information related to their models, data, and underlying infrastructure to maintain their market position.

**Mitigation: Transparency as a strategic imperative**

The history of software and technology provides us with a powerful lesson, proprietary systems can lead to slow, siloed innovation. The creation of Linux and the explosion of open source software were catalysts for an entire modern ecosystem of innovation, permitting AI to exist. Today, we see a similar trajectory in the AI ecosystem. Massive repositories like Hugging Face provide thousands of publicly available, high-quality models[24]. There has been a shift in the AI ecosystem since the early days of chatgpt, where large-scale data sets are increasingly more common[25], and even frontier model developers now release  models for research and consumption, sometimes under open source licenses[26]. These two distinct changes can be seen as an acknowledgement that community engagement, a key characteristic in open source, drives progress and adoption. Continued withholding of a model's basic architecture and training methodology isn't where the value of the model or system lies, the basic blueprint is universally available. The value has shifted to the application, refinement, and trust enabled.

While the objection that training data is intellectual property is valid, it is frequently misapplied. Truly unique and curated prioritary data sets can provide a competitive advantage and do warrant protection, however, as an industry we collectively and continuously overclassify data. The vast majority of data used in training is commonly derived from publicly available resources or represents universally known information, often through a different lens. The AI system cards are not designed to disclose key strategic components, but address consumers needs, be they models or systems, for system transparency[27] prior to buying and integrating AI enabled products within their pipelines.

---

[24] https://huggingface.co/models
[25] https://laion.ai/projects/
[26] https://openai.com/index/introducing-gpt-oss/
[27] https://kpmg.com/us/en/articles/2025/when-trust-ai-matters.html

Remember that opaqueness has a cost[28] that shows up as risk aversion and increased regulatory scrutiny, and pushes the industry back towards castle and moat security models.

**Challenge: Transparency leads to systems being exploited by bad actors**
It is a common thought that full transparency in AI system cards can unintentionally expose sensitive details about hazards and response mechanisms. Over-disclosure may give attackers a roadmap to exploit vulnerabilities or bypass safeguards, while also creating legal and reputational liabilities if stakeholders interpret documented risks as negligence or unreliability.

**Mitigation**
Early critics of open source argued that revealing code would arm attackers; in practice, security through obscurity didn't hold[29]. Openness increased scrutiny, raised the *discovery* rate of flaws, and crucially reduced time-to-fix via coordinated disclosure, CI/fuzzing, reproducible builds, and a shared CVE ecosystem, leaving software more secure overall. The lesson for AI system cards is similar: make design, evaluation methods, governance, and high-level hazards public to enable external review and accountability, while gating precise exploit paths and defense triggers such as red-team prompts, filter thresholds, bypass recipes to trusted channels. In short: open design and closed exploit details borrow the transparency that strengthened open source without giving attackers a roadmap[30].

**Challenge: Maintenance is resource intensive**
Due to some of the extra work required in keeping the system card updated, it is possible that engineering teams think of the HASC as a blocker to rapid innovation. Especially when there is uncertainty about certain fixes being properly applied or not.

**Mitigation**
Automation is the answer to reducing toil in this case. The machine readable AI system card should be designed to be generated automatically during the model build and deployment pipeline and should be available for consumers in a machine readable form. Additionally, another layer of automation can be added by agentic servers, such as MCP or others, advertising system cards via publicly available end points. Consumers, agentic or otherwise, would contact these end points, download the system card, and make decisions based on pre-set safety standards, on whether they want to use this AI system or not.

Further in this paper we discuss a JSON schema and discuss how some of the previously discussed automation can be achieved.

---

[28] https://www.oceg.org/what-does-transparency-really-mean-in-the-context-of-ai-governance/
[29] https://www.korte.co/2025/07/31/the-security-advantage-ofopen sourcee-software/
[30] https://arxiv.org/pdf/2501.18669v1

## Scenario

In this section we present a use-case scenario for an open AI system card. There may be other uses depending on what the system is and what kind of data is shared in the system card.

**The system:** A public-facing "AI Health Assistant" chatbot. Its HASC is currently at **v1.2**.

1. The incident

Live monitoring detects a spike in user reports and downvotes. An internal review discovers that users are successfully tricking the chatbot into generating plausible but dangerously incorrect medical advice by framing their questions around celebrity wellness trends.

2. Triage and hazard identification

The AI Safety team investigates and determines this is a new variant of a known hazard. They assign it a unique identifier: ASH-2025-0142: "Model safety bypass via social/pop-culture framing."

3. The fix

The engineering team implements a two-part fix:
- **Guardrail update:** They update the prompt of a separate "safety check" model, a guardrail, to be more sensitive to celebrity names and pop-culture jargon in medical queries.
- **System prompt change:** They add a line to the main chatbot's system prompt, explicitly instructing it to defer to a human medical professional when wellness trends are mentioned.

4. Updating the Hazard-Aware System Card (HASC)

After testing and deploying the fix, a new version of the HASC is published: v1.3. The key is updated in the card. Adding a new section, "Version history and changelog," would make this transparent.

---

- **v1.3 (Current) - Published 2025-07-23**

    - **Change type:** *Minor (safety enhancement)*
    - **Associated hazard(s):** *ASH-2025-0142 (Model bypass via social/pop-culture framing)*
    - **Summary of changes:** *Strengthened guardrails against medical disinformation disguised as pop-culture queries.*
    - **Affected components:**
        - *Proactive hazard analysis: Added ASH-2025-0142 to the hazard log.*

- - ■ *Embedded guardrails: Updated description of the "Medical query safety check" guardrail model to reflect its new sensitivity.*
    - ■ *System blueprint: Updated the "System prompt" documentation to include the new instruction.*
  - ○ **Linked incident report:** *[Internal link to incident report 2025-07-23a]*
- **v1.2 - Published 2025-06-15**

  - ○ **Change type:** *Minor (performance tuning)*
  - ○ **Summary of changes:** *Adjusted model parameters to reduce latency. No safety components were affected.*
- **v1.0 - Published 2025-02-01**

  - ○ **Change type:** *Major (initial release)*
  - ○ **Summary of changes:** *First public release of the AI Health Assistant system.*

## Automated generation of AI System cards

As discussed in the earlier sections of the paper, the ability to automate generation and consumption of system cards is one of the biggest factors to consider when deploying these artifacts and critical to enabling their consistent, repeatable creation within the ecosystem. In this section, we present a proposal for automated generation of system cards as part of the build process. Generating AI System cards at build time is the most accurate and efficient method of consistently capturing key data coupled with the specific model system version and benefits from existing supply chain security practices being applied such as signing. The proposal is but one example and may not be suitable for all use cases or deployment scenarios. However it should serve as a basic guide on how such automation can be implemented.

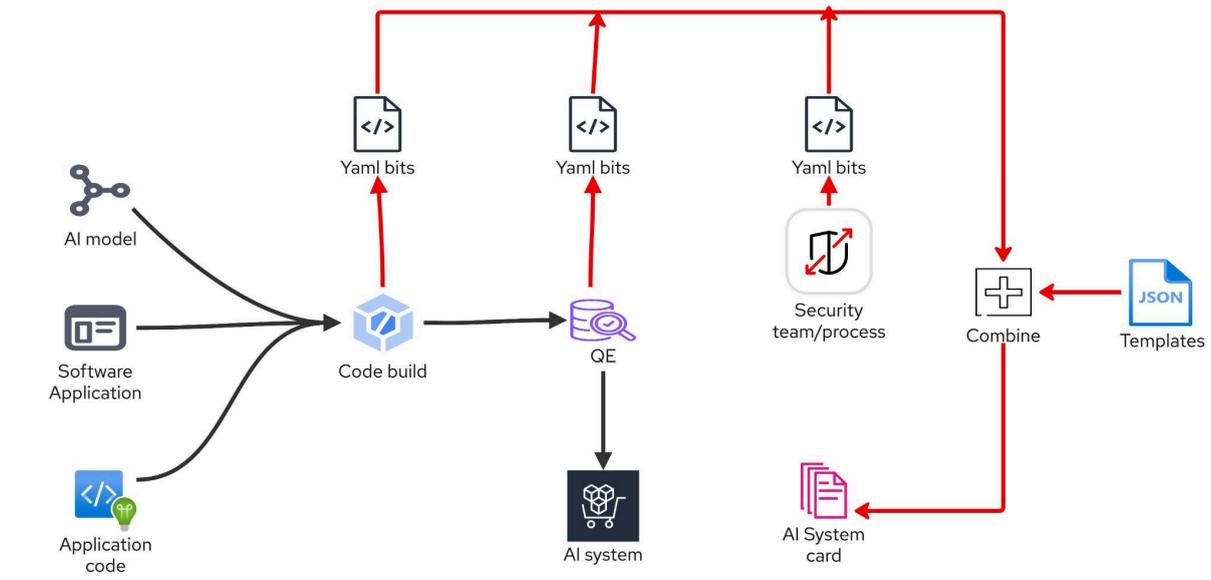

The system consists of a data source, a template, and a glue script. The glue script combines the template and the data source to produce the final artifacts. The format of the final artifacts can be varied as required and can be stored at varied places within the infrastructure.

For our proof of concept, we use yaml generated during the build process with additional inputs from Quality Engineering processes. This is optional but encouraged. Output from other teams, such as the security team is for any security or safety tracking data as the main data sources for creating the AI System card.

The yaml is generated at each stage in bits and is combined after the process is complete. The glue script combines this with a json template and creates an HTML artifact, which can be published along with the AI system.

This process allows variations, such as using a database to collect and furnish information for the system card. This can be beneficial for mature organizations collecting and storing security and other testing metadata about artifacts generated from their build system to attest what has transpired for a given software, application, or system.

The final system card can be made machine-readable only or both machine and human-readable by publishing an HTML as well as a yaml/json for machine reading. It is recommended that both are supported as consumers of an AI system will likely require the contents of the AI System card to make an informed risk decision before integration and deployment. At deployment or runtime, organizations can establish policy rules for decision or enforcement of AI systems against their risk tolerances and pre-deployment checks; this is made significantly easier when the data values used in the ruleset are machine readable.

Using automated AI System card generation, two variants of the system card can be easily created to allow internal consumption and documentation containing information, which is not suitable for external release as well as a public version.

The system card can be embedded in the final artifact of the AI system if they are to be shipped to customers. If shipping models or systems as OCI artifacts, we can embed the system card inside the OCI layers.

A proof of concept code, along with a sample json schema and yaml data, is available for review and further integration at:
https://github.com/RedHatProductSecurity/ai-system-card

## AI system cards and ISO/IEC 42001 standard

The Hazard-Aware System Card (HASC) and the ISO/IEC 42001:2023[31] standard aim to promote responsible AI, but they approach this goal from different perspectives. The HASC is presented as a dynamic, transparent, and machine-readable documentation framework for individual AI systems, focusing on their security and safety posture throughout their lifecycle. In contrast, ISO/IEC 42001 is an international standard for an AI Management System (AIMS), providing requirements and guidance for organizations to responsibly develop, provide, or use AI systems by establishing, implementing, maintaining, and continually improving a comprehensive management system.

The system card, specifically the HASC framework, directly supports and aligns with several core tenets of the ISO/IEC 42001 standard. For example, the HASC's four categories of content – System Overview and Intended Use, Model and Data details, Evaluation and Performance metrics, and Limitations and Known Biases, provide crucial information that can fulfill various ISO requirements. The "System Overview and Intended Use" section of a system card offers governance and contextual mapping necessary for ISO's risk classification and defining policies and controls, especially concerning inappropriate use. Similarly, the "Limitations and Known Biases" section directly guides and informs the AI system impact assessments required by ISO/IEC 42001, Clause 6.1.4 and Annex A.5. The HASC's emphasis on documenting data provenance and pedigree, along with evaluation and performance metrics, aligns with ISO's detailed requirements for data for AI systems, Annex A.7, and AI system verification and validation, Annex A.6.2.4.

Ideally generated automatically within the software delivery pipeline as a machine-readable "living document," the HASC's design directly addresses ISO/IEC 42001's pervasive requirement for "documented information," Clause 7.5. This automated and standardized approach allows better version control, auditable artifacts, and supports continual improvement by recording provenance, updates, and remediated hazards across releases.

---

[31] https://www.iso.org/standard/42001

The proposed AI Safety Hazard (ASH) ID within the HASC framework provides a structured way to track and communicate safety flaws, complementing the incident response and hazard remediation section of the card. This is essential for an ISO-compliant organization's AI risk treatment, Clause 6.1.3 and Annex A.6.2.6 related to repairs and updates, and communication of incidents, Annex A.8.4. While the HASC is a specific artifact for individual AI systems, ISO/IEC 42001 provides the overarching framework for an organization to manage *all* of its AI systems responsibly, meaning that well-implemented HASCs would serve as valuable evidence and tools within an ISO-certified AIMS.

## Industry engagement and further directions

To realize the full potential of the Hazard-Aware System Card (HASC) framework, a concerted effort is required to engage with industry partners, open source communities, and regulatory bodies. Authors suggest focusing on building consensus and extending the HASC's capabilities beyond its initial scope.

- **Standardization with industry bodies:** Collaboration is required within the AI community and the AI industry. The goal is to develop a standardized, machine-readable schema for system cards that can be adopted across the industry, [32]addressing the current lack of comparability and consistent fields.
- **Establish a shared hazard ecosystem:** Similar to the Common Vulnerabilities and Exposures (CVE) system for software, we will work to promote the use of a common AI Safety Hazard (ASH) identifier. This will allow for a shared, public database of known AI hazards and their remediations, fostering collective defense and a faster time-to-fix for the entire ecosystem. We proposed such a system in the last paper[33].
- **Encourage phased transparency:** Recognizing the competitive pressures and the "magic sauce" dilemma, we will advocate for a tiered, phased approach to transparency. This strategy allows companies to gradually increase their disclosure, building trust with consumers who are increasingly looking for transparent systems before integrating them into their pipelines.

## Suggestions for moving forward

- **Automation and tooling integration:** The long-term viability of the HASC depends on its integration into development workflows. Future work will focus on developing automated tooling to generate HASC data directly from model build and deployment pipelines. This includes creating APIs for automated system card generation and consumption, making it a "living document" that reflects real-time changes.
- **Integration with regulatory frameworks:** We will explore how HASC can serve as a compliance vehicle for emerging AI regulations, such as the EU AI Act and NIST AI RMF. The structured, auditable nature of the HASC's version history and incident log could provide a clear record of due diligence, helping organizations demonstrate responsible development and risk management to regulators.

---

[32] https://github.com/RedHatProductSecurity/ai-system-card
[33] https://arxiv.org/pdf/2411.12275

## Closing remarks

The Hazard-Aware System Card (HASC) framework emerges as a vital blueprint for the future of AI governance, addressing the critical gap between powerful AI models and the complex systems in which they are deployed. By expanding upon traditional model and system cards, this framework introduces a dynamic, living document that not only records an AI system's architecture and intent but also documents a proactive analysis of potential hazards.

The proposal for a standardized AI Safety Hazard (ASH) ID, akin to a CVE, represents a significant step toward creating a shared language for communicating safety flaws and their remediations across the industry. Ultimately, the HASC aims to serve as a central source of truth, empowering developers and external stakeholders to make informed decisions and fostering a new era of end-to-end transparency and accountability in AI development.

The success of the HASC framework hinges on its broad adoption and integration into the broader AI ecosystem, and this paper presents several strategies to overcome key challenges. The shift from security-by-obscurity to transparency, as demonstrated by the open source community, can serve as a model for AI safety.

This framework is not merely a technical proposal but a call to action for the industry to collectively build a foundation of trust and reliability, ensuring that AI's transformative power is harnessed safely for the benefit of all.

Mitchell, M., et al. (2018). Model Cards for Model Reporting.
https://arxiv.org/abs/1810.03993

Meta AI. (n.d.). System-Level Transparency of Machine Learning.
https://ai.meta.com/research/publications/system-level-transparency-of-machine-learning

Korte, S. (2025). Open Design, Closed Exploit Details. https://arxiv.org/pdf/2501.18669v1

## Definitions

**AI model:** An AI model is a software framework that learns from data to make predictions or decisions, simulating aspects of human cognitive processes. *Source: GENAI Commons*[34]

**AI Security**: The field of security focused on protecting and securing AI, AI Systems, and AI workloads. It covers the security of: data and the model was trained on, the supply chain of the AI Model, security capabilities that support the operational security of models such as prompt injection protection and data exfiltration detection, and other related areas.

**AI Safety**: The field of study and practices to ensure AI systems operate in a manner that is safe and aligned with human values, preventing harm to individuals and society. It is considered a subset of AI Trustworthiness. *Source: Modified for Red Hat from GENAI Commons*

**AI system**: LLMs and everything support libraries and software, for example pytorch + inference software. RHEL AI is considered an AI system, as it is packaged to include the necessary components and software for use.

**Embargoed security flaws:** Embargoed security flaws are vulnerabilities or weaknesses in software, hardware, or systems that have been identified but are not yet publicly disclosed. Information about these flaws is temporarily restricted to a limited group of trusted parties—such as developers, security teams, and affected vendors—under an agreement not to share details externally until a specified date or event. The purpose of the embargo is to allow time for patches or fixes to be developed, tested, and distributed before the vulnerability becomes widely known.

**LLM Guardrails:** Guardrails are the set of security and safety controls that monitor and dictate a user's interaction with a LLM application. They are a set of programmable, rule-based systems that sit in between users and foundational models in order to make sure the AI model is operating between defined principles in an organization.

**Model Cards:** A model card is a type of documentation that is created for, and provided with, machine learning models. A model card functions as a type of data sheet, similar in principle to the consumer safety labels, food nutritional labels, a material safety data sheet or product spec sheets.

---

[34] https://genaicommons.org/glossary/ai-model/

**Security Weakness:** A weakness is a condition in a software, firmware, hardware, or service component that, under the right circumstances, could contribute to the introduction of vulnerabilities.

**Safety Hazard**: An unexpected behavior or output outside of the defined intent and scope of a system's or software's design. Safety hazards are linked with producing harmful content or outputs that can cause social, economic, and environmental harm to consumers and users.

**Sustainable AI**: the development and use of carbon neutral and carbon negative practices to minimize the negative environmental impacts of AI technologies. While considered its own independent domain, the ecological and corresponding human harm that results from unsustainable AI development and use aligns with Responsible AI – directly influencing the duration of benefits for humans against environmental realities (akin to corporate social responsibility focuses). It is related to AI Ethics.